\documentclass[aps,prl,showpacs,superscriptaddress,floatfix,amssymb,amsmath]{revtex4}
\usepackage{graphicx}
\begin{document}

\title{NMR Time Reversal as a Probe of Incipient Turbulent Spin Dynamics}
 
\author{M.~E.~Hayden}
\affiliation{Physics Department, Simon Fraser University, 8888
University Drive, Burnaby BC, Canada V5A 1S6}
\author{E.~Baudin}
\author{G. Tastevin}
\author{P.J.~Nacher}
\affiliation{Laboratoire Kastler Brossel, Ecole Normale
Sup{\'e}rieure; CNRS; UPMC; 24 rue Lhomond, F75005 Paris, France}

\begin{abstract}
We demonstrate time reversal of nuclear spin dynamics in highly
magnetized dilute liquid $^3$He-$^4$He mixtures through effective
inversion of long-range dipolar interactions. These experiments,
which involve using magic sandwich NMR pulse sequences to generate
spin echoes, probe the spatiotemporal development of turbulent spin
dynamics and promise to serve as a versatile tool for the study and
control of dynamic magnetization instabilities. We also show that a
repeated magic sandwich pulse sequence can be used to
dynamically stabilize modes of nuclear precession that are otherwise
intrinsically unstable. To date, we have extended the effective
precession lifetimes of our magnetized samples by more than three
orders of magnitude.
\end{abstract}

\pacs{82.56.Jn, 75.50.Mm, 76.60.Jx, 76.60.Lz \\ \\
Published in {\it  Phys. Rev. Lett. 99, 137602 (2007)}\\
http://link.aps.org/abstract/PRL/v99/e137602}
\maketitle

Elementary treatments of nuclear magnetic resonance (NMR) ignore
collective effects. Individual spin dynamics are assumed to be
independent of the sample magnetization $\mathbf{M}$, and are thus
governed by linear differential equations. This approximation is
justified for many -- but certainly not all -- practical
applications. One of the most well known and pervasive
counter examples is the phenomenon of radiation damping
\cite{vlassenbroek95}, in which the emf induced in the pickup coil
by $\mathbf{M}$ drives a current; the magnetic field associated with
this current in turn acts on $\mathbf{M}$, driving it into alignment
with the static field $\mathbf{B}_0$ in a nonlinear manner. A less
common but more insidious problem arises when the magnetic field
produced by the magnetized sample becomes large enough to directly
influence spin dynamics
\cite{candela94,jeener99,sauer01,ledbetter02,jeener02}. This leads
to a rich variety of nonlinear effects that range from spectral
clustering to precession instabilities and spin turbulence. The
former arises when small-angle NMR tipping pulses are applied to the
sample (generating small angular displacements of $\mathbf{M}$ from
$\mathbf{B}_{\rm 0}$), and is manifest by the spontaneous appearance
of long-lived geometry-dependent modes of coherent nuclear
precession. The latter occurs when large-angle tipping pulses are
employed, and involves an exponential growth in the complexity of
spatially inhomogeneous magnetization patterns. Far from being mere
curiosities, phenomena arising from nonlocal interactions
(including the joint action of `distant dipolar fields' and
radiation damping \cite{lin00,datta06}) threaten to limit (or
profoundly alter) the operation of state-of-the-art high-field
high-resolution NMR spectrometers. They are equally important for
understanding the dynamics of polarized quantum fluids including
Bose-Einstein condensates \cite{giovanazzi02}, superfluid $^3$He
\cite{borovik84,corruccini78}, degenerate $^3$He-$^4$He mixtures
\cite{perisanu06}, and two-dimensional H gases \cite{vasilyev02}.
From yet another perspective, the collective effects induced by
distant dipolar fields can be used to amplify weak spin precession
signals \cite{ledbetter05} and to enhance the sensitivity of
precision searches for CP-violating permanent electric dipole
moments \cite{ledbetter02}.

Here we describe NMR time reversal experiments that -- for the first
time -- shed light on the extent to which the deleterious effects of
spin turbulence can be controlled or suppressed. Moreover, they lay
the foundation for a systematic and quantitative investigation of
the onset of dynamical magnetization instabilities under diverse
conditions. Our experiments are performed on dilute non-degenerate
mixtures of hyperpolarized $^3$He in liquid $^4$He. They are based
on the use of magic sandwich (MS) pulse sequences
\cite{rhim71,hafner96}: continuous bursts of tipping pulses that are
appropriately timed and phased so as to refocus spin dynamics
associated with dipolar interactions, leading to the formation of
echoes. The MS pulse cycle and numerous variants are indispensable
tools in the field of solid-state NMR but, to the best of our
knowledge, have not been used previously for liquid-state NMR
\cite{caution}.

Our samples are confined to a slightly-prolate 0.44~cm$^3$
spheroidal Pyrex cell treated with Cs metal to suppress wall
relaxation \cite{note,piegay02}, and are maintained at temperatures
$T \sim 1$~K. The sample cell communicates with a room temperature
metastability-exchange optical pumping cell and various gas
reservoirs via a narrow Pyrex tube. The entire apparatus is immersed
in a 2.3~mT magnetic field shimmed to $\pm$20~ppm over the sample
cell and actively stabilized to $\pm$10~ppm against fluctuations in
the background laboratory field. We control the volume of liquid
admitted to the cell (typically 40~\% full), the concentration $X$
of $^3$He in the mixture (typically 1-5~\%), and the nuclear
polarization (typically~$\lesssim 40$~\%). We can thus manipulate
the shape of the sample, the initial amplitude of its magnetization
$M$, and the $^3$He diffusion coefficient $D$. Note that it is
convenient to report $M$ in terms of a characteristic dipolar field
$B_{\rm dip}=\mu_0 M$ or dipolar frequency $F_{\rm dip}=\gamma
B_{\rm dip}/2\pi$, where $\mu_0$ is the permeability of free space
and $\gamma$ is the $^3$He gyromagnetic ratio.

Our pulsed NMR experiments are performed at 74~kHz using orthogonal
transmit and receive coils. The transmit coils are actively-shielded
\cite{bidinosti05} to suppress the induction of eddy currents in
nearby conducting structures. This enables us to apply the intense
and sustained rf (i.e. $\mathbf{B}_1$) fields that are necessary for
NMR time reversal experiments without causing the sample to warm. It
also ensures that eddy currents do not distort the applied rf field,
which is uniform to within $\pm 200$~ppm over the sample cell. The
receive coils are tuned in a tank configuration with a quality
factor $Q_0\sim10$. The filling factor $\eta$ is of order $0.02$
when the cell is completely full, and an externally adjustable
`$Q$-spoiling' resistor is used to ensure that radiation damping
fields are small compared to those generated by the magnetized
sample. Typically a value $Q \sim 2$ was employed for the work
presented here.


\begin{figure}[h]
\includegraphics[width=10cm]{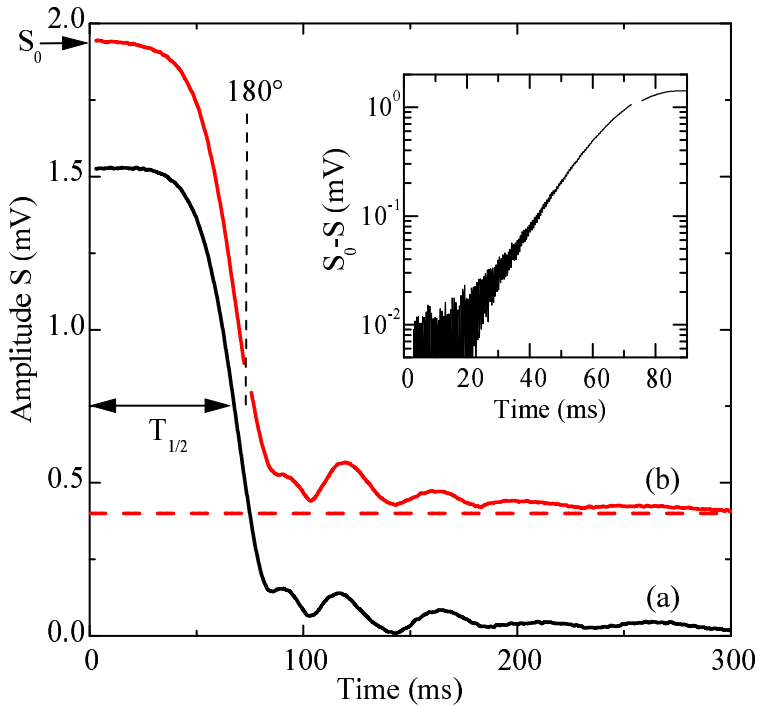}
\caption{[color online] FID response of polarized $^3$He in liquid
$^4$He (a) to a $90^\circ$ rf tipping pulse and (b) to a
$90^\circ$-$\tau$-$180^\circ$ sequence. The time at which the FID
amplitude $S$ drops to half of its initial value $S_0$ is denoted
$T_{\scriptscriptstyle 1/2}$, which is of order 70~ms for these
examples. Trace (b) has been displaced upward by 0.4~mV for the sake
of clarity. The inset shows the deviation of trace (b) from its
initial amplitude. These data were acquired at $T=1.14$~K, with
$X\sim 2$~\% and $B_{\rm dip}\sim0.9$~$\mu$T for trace~(a) and
$X\sim 5$~\% and $B_{\rm dip}=0.7$~$\mu$T for trace~(b).
\label{EchoFdip}}
\end{figure}

Figure \ref{EchoFdip} summarizes data from two experiments that
illustrate some of the unusual behavior exhibited by highly
magnetized liquids. The first experiment involves application of a
single $90^\circ$ rf tipping pulse. The corresponding free induction
decay (FID) is shown as trace (a). The signal amplitude $S$ is
roughly constant for several tens of ms, during which time the
precessing transverse nuclear magnetization is necessarily uniform.
This is followed by a precipitous drop that culminates in a number
of small revivals. Similar experiments performed over a range of
initial conditions indicate that the characteristic time scale for
the abrupt decrease in signal amplitude (i.e. $T_{\scriptscriptstyle
1/2}$) is inversely proportional to $B_{\rm dip}$ \cite{nacher02}.

Evidence that this behavior is not dominated by terms in the
equation of motion linear in $\mathbf{M}$ is provided by the second
experiment, in which the initial 90$^\circ$ pulse is followed
$70$~ms later by a 180$^\circ$ pulse. Comparison of traces (a) and
(b) reveals that the 180$^\circ$ pulse has little if any effect on
the FID envelope. In particular, there is no obvious sign of a Hahn
(or inhomogeneous) spin echo $140$~ms after the start of the
experiment. Closer examination of these data shows that the
deviation of the FID amplitude $S$ from its initial value $S_0$
grows exponentially with time $t$ (inset to Fig. \ref{EchoFdip}):
i.e. $S_0-S \sim\exp(2\Gamma t)$. This behavior is expected for an
infinite spin system dominated by distant dipolar interactions
\cite{jeener02}, and has been interpreted as the onset of spin
turbulence \cite{jeener99}. We have performed exact numerical
simulations of these experiments for small numbers of dipolar-coupled spins diffusing on a cubic lattice (up to
$32\times32\times32$ sites) with periodic boundary conditions
\cite{nacher02}, and observe excellent agreement between measured
and calculated growth rates. In both cases $\Gamma/F_{\rm dip}\sim
2.3$, which is somewhat smaller than the value
$2\sqrt{2}\pi/3\sim3.0$ predicted for an infinite spin system
\cite{jeener02}.

The experimental challenge presented by these data is to devise a
method for probing the spatial and temporal evolution of the nuclear
magnetization before, during, and {\em after} the point at which it
spontaneously and abruptly averages to zero in Fig.~\ref{EchoFdip}.
The approach we explore here as a first step toward this goal is the
application of MS pulse sequences to reverse the effective time
evolution of dynamics associated with dipolar interactions, and
thereby attempt to generate spin echoes. Thorough descriptions of
the manner in which time reversal is accomplished can be found in
the literature \cite{rhim71}. For our purposes the essential
features of these multiple-pulse sequences are that (a) the applied
rf field must be large compared to the dipolar fields and (b) the
effective rate at which the dipolar-coupled system evolves in time
as the rf fields are applied is scaled by a factor of $-1/2$.

\begin{figure}[h]
\includegraphics[width=10cm]{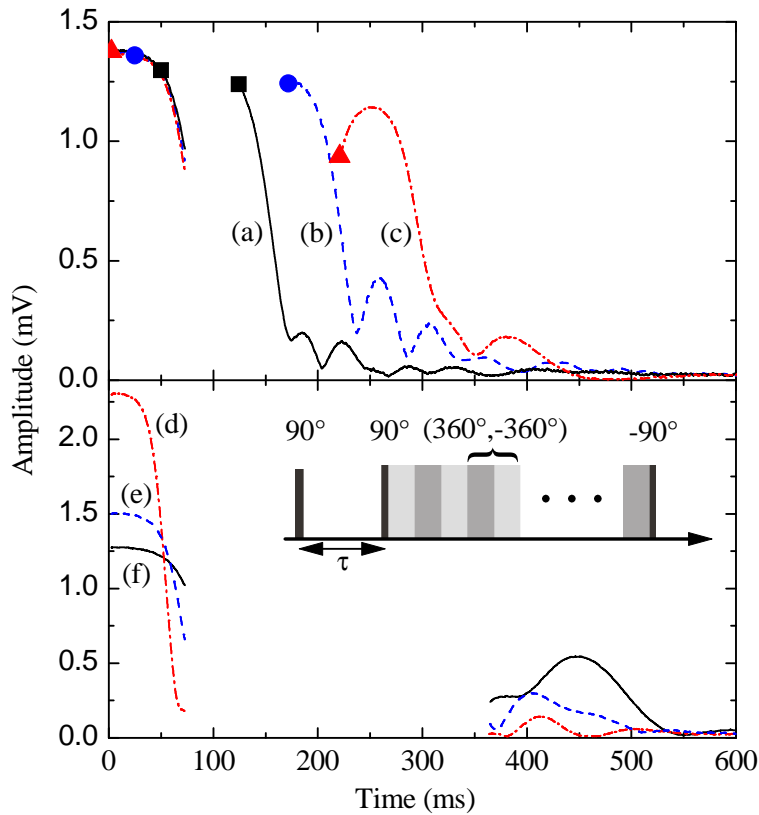}
\caption{[color online] Spin echo formation: A $90^\circ$ pulse is
applied at $t=0$, after which the magnetization is allowed to evolve
for $\tau=70$~ms. MS sequences of various durations are then used to
progressively refocus the magnetization. For traces (a) - (c), the
amplitude of the FID immediately after the MS probes the rotary echo
formed in the doubly-rotating frame. The symbols {\tiny
{$\blacksquare$}}, $\bullet$, {\scriptsize $\blacktriangle$}
indicate pairings between points on the initial FID to which the
state of the magnetization is nominally returned and the ensuing
free precession signal. For traces (d) - (f), the amplitude of the
FID probes the magic sandwich echo formed in the rotating frame
after the magnetization is first `unwound' back to $t=-\tau$ before
being `released' and allowed to undergo further free evolution.
These data were acquired under conditions similar to those outlined
in Fig. \ref{EchoFdip}. Inset: A classic magic sandwich
\cite{rhim71} consists of a continuous train of
$(180^\circ_x,-180^\circ_x)$ rotation pairs sandwiched between a
leading $90^\circ_y$ and a trailing $-90^\circ_y$ rotation. We have
used $360^\circ$ (rather than $180^\circ$) rotations to reduce the
accumulation of phase errors associated with offsets between the rf
and Larmor frequencies. Each $360^\circ$ rotation is accomplished in
2~ms, corresponding to $B_1=15$~ $\mu$T. Note that for trace (c) the
sense of the terminal $90^\circ$ rotation was reversed (i.e.
$90^\circ_y$ instead of $-90^\circ_y$) in order to also refocus
interactions with a linear dependence on $\mathbf{M}$. This does not
significantly alter the initial amplitude of the echo or its
subsequent growth. \label{MagicS}}
\end{figure}

Figure~\ref{MagicS} shows data from several experiments that are
analogous to the ineffective $90^\circ$-$\tau$-$180^\circ$ sequence
of Fig.~\ref{EchoFdip} except for the fact that the $180^\circ$
tipping pulse has been replaced with MS sequences of various
durations. The rf used for these sequences is typically adjusted to
be within 1~Hz of the $^3$He Larmor frequency. The most striking and
important feature of these data is that large NMR signals are
detected long after the net transverse magnetization normally would
have averaged to zero. The upper half of Fig. \ref{MagicS} shows
data from three experiments that progressively `unwind' the state of
the dipolar-coupled system back to times $t=2\tau/3$, $t=\tau/3$,
and $t=0$, as indicated by the paired-symbols {\tiny
\raisebox{0.03cm}{$\blacksquare$}}, $\bullet$, and {\scriptsize
\raisebox{0.03cm}{$\blacktriangle$}}. The initial amplitudes of
these echoes nominally probe the amplitude of the `rotary echo'
\cite{rhim71,rhim72} that is formed in the doubly-rotating frame
\cite{redfield55} during the MS sequence. In the first case -- trace
(a) -- the magnitude of the transverse magnetization is returned to
a value that is close to that which is anticipated. Moreover, the
shape of the ensuing FID mimics the original decay (cf.
Fig.~\ref{EchoFdip}). In the last case -- trace (c) -- the amplitude
of the signal following the MS starts out noticeably smaller than
anticipated but then unexpectedly grows rather than decaying.
Clearly, the deficit in the initial echo amplitude is not associated
with an irreversible process. The shape of the FID labeled (b) is
qualitatively intermediate between traces (a) and (c). The lower
part of Fig.~\ref{MagicS} shows data from three experiments that
nominally `unwind' the state of the dipolar-coupled system back to a
time $t=-\tau$, resulting in the formation of a `magic sandwich
echo' in the rotating frame \cite{rhim71} after a further free
evolution period of duration $\tau$. Increasing the initial
magnetization increases the rate at which magnetization patterns
develop, and so this series of experiments effectively probes the
sample at successively later stages of evolution.

It is not surprising that we observe imperfect refocussing of the
transverse nuclear magnetization in these experiments. Irreversible
processes (e.g. diffusion) necessarily compete with the
spatiotemporal evolution of the magnetization and progressively
attenuate structures on larger and larger scales as time progresses.
At the same time, while the MS tends to refocus the family of
magnetization patterns that develop during the initial FID, one
expects a complementary family of instabilities \cite{jeener02} to
be established and grow during the time reversal sequence. Technical
imperfections in the time reversal sequence will also contribute to
echo attenuation. A detailed and systematic study of these rich and
complex effects is required before quantitative statements can be
made regarding the onset and development of spin turbulence.

\begin{figure}[h]
\includegraphics[width=10cm]{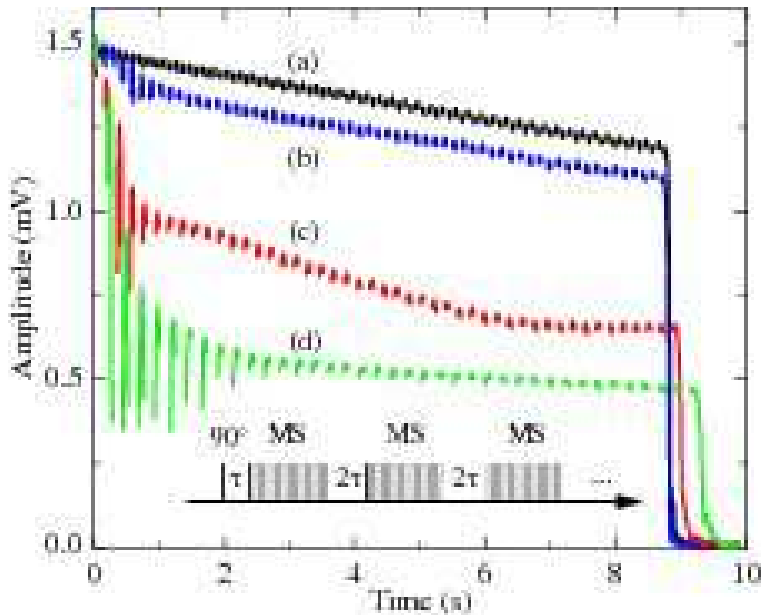}
\caption{[color online] Spin echo trains: Traces (a) - (d)
correspond to echo trains that are observed when RMS sequences with
delay times $\tau = 16$, 24, 32, and 40~ms, respectively, are
applied to the magnetized liquid. The MS cycles used to obtain these
data are identical to those described in the caption of Fig.
\ref{MagicS} except for the facts that each one terminates with an
additional $180^\circ_y$ rotation (cf. trace (c) of
Fig.~\ref{MagicS}) to refocus interactions linear in $\mathbf{M}$,
and that the symmetry of alternate MS cycles is inverted to reduce
the accumulation of phase errors. That is, odd MS cycles are of the
form [$90^\circ_y$, $n\times(360^\circ_x$, $-360^\circ_x)$,
$90^\circ_y$] while even MS cycles are of the form [$-90^\circ_y$,
$n\times(-360^\circ_x$, $360^\circ_x)$, $-90^\circ_y$]. These data
were acquired at $T \sim 1.1$~K, with $X\sim 4.5$~\% and $B_{\rm
dip} \sim 0.8$~$\mu$T. In the limit $B_{\rm dip}\rightarrow 0$,
CPMG-based measurements yield $D \sim 2\times 10^{-7}$~m$^2$/s.
\label{MagicT}}
\end{figure}

The fact that spin echoes {\em can} be generated in highly
magnetized liquids leads one to contemplate the use of more
sophisticated NMR tools. To illustrate this point, we describe here
a simple but significant extension of the basic MS sequence that
clearly merits further investigation. A conventional CPMG sequence
($90^\circ$-$\tau$-$180^\circ$-$2\tau$-$180^\circ$-$2\tau \ldots$)
acts on a system dominated by interactions that are linear in
$\mathbf{M}$, repeatedly inverting the phase of the precessing
magnetization to produce an echo train. By analogy, a Repeated Magic
Sandwich (RMS) sequence consisting of a train of MS cycles
interspersed with appropriate delays should likewise act on a system
dominated by distant dipolar interactions to form an echo train. The
validity of this conjecture is borne out by the data shown in
Fig.~\ref{MagicT}. The RMS sequence used to generate these data is
of the form ($90^\circ$-$\tau$-${\rm MS}$-$2\tau$-${\rm MS}$-$2\tau
\ldots$), where the duration of each MS cycle is $4\tau$ (i.e. long
enough to effectively `reverse' the time evolution of the dipolar
coupled system for a time $2\tau$). The four traces correspond to
experiments performed with different delay times $\tau$. In each
case, the RMS sequence is applied for $\sim 9$~s, at which point
$\mathbf{M}$ is allowed to evolve without further interruption.

Figure \ref{MagicT} reveals two distinct phases in the evolution of
the magnetization during the application of RMS sequences. Each echo
train starts with a period that we associate with imperfect
refocussing of the magnetization. That is, behavior similar to the
examples presented in Fig. \ref{MagicS}. Increasing the length of
time during which the magnetization is allowed to evolve between
successive magic sandwiches results in greater losses. This initial
period is followed by a second phase during which magnetization
losses continue to be incurred, but at significantly lower rates.
Comparison with Figs.~\ref{EchoFdip} and~\ref{MagicS} reveals that
the apparent lifetimes of these dynamically sustained precession
signals are in some cases more than 3 orders of magnitude longer
than the characteristic coherence time (e.g. $T_{\scriptscriptstyle
1/2}$) of the freely evolving magnetization. Exact numerical
simulations of these experiments (analogous to those described
above) have been used to investigate the influence of diffusion and
rf field amplitude. They reveal similar -- but not identical --
behavior. The initial period during which losses are incurred is
observed, as is the transition to a second evolutionary phase and
the dependence of this transition on both $\tau$ and $B_{\rm dip}$.
On the other hand, our model calculations yield a steady state
solution for which we are not able to resolve a decay rate: that is,
a plateau. This suggests that our experimentally-determined echo
trains may yet be limited by systematic rather than intrinsic
effects. One possibility is the small but finite heat load placed on
our cryostat by eddy currents induced during RMS sequences, which
would in turn lead to evaporation of $^3$He atoms. Another
possibility is imperfect refocussing of the magnetization caused by
imperfect phasing of rf tipping pulses. Finally, it is worth
commenting on the terminal FID at the end of each RMS sequence.
Analysis of these decays yields values of $\Gamma$ that are
proportional to the terminal echo amplitude and consistent with
those observed following a single $90^\circ$ tipping pulse (cf. Fig.
\ref{EchoFdip}).  This strongly suggests that the precessing
magnetization is highly uniform at the peak of each revival, and
that the `seeds' from which instabilities ultimately develop
\cite{jeener02} are small.

In summary, we have demonstrated the production of NMR spin echoes
and spin echo trains in magnetized liquids under conditions where
nuclear spin dynamics (in the rotating frame) are dominated by
nonlinear and nonlocal (or distant) dipolar interactions. Our
experiments, which are based on the use of magic sandwich time
reversal pulse sequences, probe the rich spatiotemporal evolution of
the nuclear magnetization distribution that results from competition
between nonlinear spin dynamics and diffusion. They promise to
yield quantitative information about the progression and extent of
disorder, as well as new methods to control or suppress the onset of
instabilities.


\begin{acknowledgments}
MH gratefully acknowledges support from the CNRS (France), the ENS
(Paris), and NSERC (Canada).
\end{acknowledgments}


\begin{thebibliography}{99}
\bibitem{note} Wall contributions to the NMR relaxation times $T_1$
and $T_2$ are of order several hours under these conditions.
\end{thebibliography}


%



\end{document}